\documentclass{article}
\usepackage{arxiv}

\usepackage[utf8]{inputenc}
\usepackage[T1]{fontenc}

\usepackage{authblk}

\usepackage{hyperref}       
\usepackage{url}            
\usepackage{booktabs}       
\usepackage{amsfonts}       
\usepackage{nicefrac}       
\usepackage{microtype}      

\usepackage{graphicx}
\usepackage{natbib}
\usepackage{doi}

\title{Acoustic characterization of speech rhythm: going beyond metrics with recurrent neural networks}

\date{}

\renewcommand{\headeright}{pre-print}
\renewcommand{\undertitle}{}

\renewcommand{\shorttitle}{Acoustic characterization of speech rhythm: going beyond metrics with recurrent neural networks}

\hypersetup{
pdftitle={Acoustic characterization of speech rhythm: going beyond metrics with recurrent neural networks},
pdfauthor={François Deloche, Laurent-Bonnasse Gahot, Judit Gervain},
pdfkeywords={speech rhythm, deep learning, rhythm metrics},
}

 \author[1, 2, *]{François Deloche}
 \author[3]{Laurent Bonnasse-Gahot}
 \author[4, 5, 6]{Judit Gervain}

 \affil[1]{Department of Speech, Language, and Hearing Sciences, Purdue University, West Lafayette, USA}
 \affil[2]{Hearing Technology Lab, Department of Information Technology, Ghent University, Ghent, Belgium}

 \affil[3]{Centre d’Analyse et de Mathématique Sociales, CNRS, EHESS, Paris, France}

 \affil[4]{Integrative Neuroscience and Cognition Center, CNRS \& Université Paris Cité, Paris, France}

 \affil[5]{Department of Developmental and Social Psychology, University of Padua, Padua, Italy}

 \affil[6]{Padova Neuroscience Center, University of Padua,
 Padua, Italy}

\affil[*]{e-mail: \texttt{francois.deloche@polytechnique.org }}

\usepackage{amssymb,amsmath}

\begin{document}

\maketitle

\begin{abstract}
Languages have long been described according to their perceived rhythmic
attributes. The associated typologies are of interest in
psycholinguistics as they partly predict newborns' abilities to
discriminate between languages and provide insights into how adult listeners process non-native languages. Despite the relative success of rhythm metrics in supporting the existence of linguistic rhythmic classes, quantitative studies have yet to capture the full complexity of 
temporal regularities associated with speech rhythm. We argue that deep learning offers a powerful pattern-recognition approach to advance the characterization of the acoustic bases of speech rhythm. To explore this hypothesis, we trained a medium-sized 
recurrent neural network on a language identification task over a large
database of speech recordings in 21 languages. The network had access to the amplitude envelopes and a variable identifying the voiced segments, assuming that this signal would poorly convey phonetic information but preserve prosodic features. The network was able to identify the
language of 10-second recordings in 40\% of the cases, and the language
was in the top-3 guesses in two-thirds of the cases.
Visualization methods show that representations built from the network
activations are consistent with speech rhythm typologies, although the
resulting maps are more complex than two separated clusters between
stress and syllable-timed languages. We further analyzed the model by identifying correlations between network activations and known speech rhythm metrics. The findings illustrate the potential of deep
learning tools to advance our understanding of speech rhythm through the identification and exploration of linguistically relevant acoustic feature spaces.
\end{abstract}

\section{Introduction}\label{introduction}

When listening to different languages, even ones we do not know or understand, we often have the impression that some languages, like Spanish and Italian, have similar rhythms, while others, like Japanese and English, are quite distinct in their rhythmicity. The percept of linguistic rhythm is so powerful that newborn infants can rely on it to distinguish rhythmically different languages \citep{Nazzi1998, Mehler2000}, and adults use the rhythmic template of their native language when listening to foreign languages \citep{Cutler1994}. Yet, the acoustic correlates of linguistic rhythm in the speech signal are only partially understood. Theories of speech rhythm originally relied on the isochrony principle,
 which holds that speech is organized into units of equal duration \citep{Pike1945, Abercrombrie1967}. A dichotomy between `syllable-timed' and `stress-timed' languages was postulated, depending on whether the isochronous units were believed to be the syllables or the intervals between stressed syllables. The isochrony principle turned out not to be supported by empirical evidence \citep{Roach1982, Dauer1983} and it is now discarded, but the term `speech rhythm' and the
associated typology of languages have remained in use. Despite the 
absence of isochronous units, the systematic alternations of strong and weak elements at different levels of the phonological hierarchy gives rise to a sense of rhythm \citep{Langus2017}. At the physical level, several dimensions are involved  and contribute to the perception
of prominence, mainly intensity, duration, pitch, and vowel quality
\citep{Terken2000}. Research on speech rhythm has thus moved from the
search of simple isochronous patterns to
the more challenging task of characterizing potentially weaker
regularities affecting a larger number of dimensions
\citep{Bertinetto1989,Kohler2009a, Cumming2010, Turk2013}. One fruitful approach, originally proposed by Dauer
\citep{Dauer1983}, has been to highlight correlations between the rhythmic
classes and several phonological differences between languages, such as the complexity of
 syllable structure or the presence of reduced vowels. This description motivated the development of speech rhythm metrics defined
by summary statistics on the duration of consonantal or vocalic
intervals \citep{ramus1999a, Grabe2002}, which were among the first studies to provide quantitative evidence for the existence of rhythmic classes. 
These studies also coincided with a renewed interest in speech rhythm within the psycholinguistics community, following the observation that newborns can discriminate between languages belonging to different rhythmic classes
\citep{Moon1993, Nazzi1998, Mehler2000}, even if those languages were unfamiliar to them, i.e. not spoken by their mothers during pregnancy. This ability allows newborn infants to discover that there are multiple languages in their environments, if they are born multilingual between rhythmically different languages  \citep{gervain10}.  Similarly, adults have been shown to rely on the rhythm of their native language even when listening to foreign languages \citep{Cutler1994}. 
Developing a more quantitative foundation for speech rhythm is also a goal in applied areas like Computer Assisted Language Learning, where it could for example enhance the assessment of non-native speakers' oral proficiency \citep{Kyriakopoulos2019}. 

Despite the relative success of speech rhythm metrics in separating prototypical syllable- and stress-timed
languages, they have been criticized for being unreliable, since variations induced by speech rate, speaker identity or speech
corpus within the same language can exceed cross-linguistic variation
\citep{Arvaniti2009, Wiget2010, Arvaniti2012}. Some authors even argue that research based on rhythm metrics 
 has reached its limit \citep{Rathcke2015} and has not  sufficiently characterized the timing regularities of speech \citep{Arvaniti2009}.

These shortcomings call for new tools to revisit the acoustic bases
of speech rhythm. These tools, as we argue in this paper, could come from the field of deep learning. 
This approach aligns well with a trend that has emerged over the last decade in cognitive science, characterized by the increasing use of deep neural networks (DNNs) in the study of sensory systems and perception \citep{Yamins2016, Kell2019, Storrs2019}.

DNNs are compositional models made of successive layers of linear and
non-linear transformations. They contain a large number of variables
that define the `synaptic' weights of the model units called artificial
neurons. When a DNN is trained on a large database of examples, it can
learn a complex non-linear relationship between input data and outputs
(labels) \citep{LeCun2015}. DNN models trained in a supervised fashion can achieve
human-level performance on tasks that individuals perform in their
everyday environment such as visual object recognition \citep{Krizhevsky2017} or
speech recognition \citep{Yu2015}. 


To test whether a deep learning model could provide insights into cross-linguistic aspects of speech rhythm, we trained a recurrent neural network with long short-term
memory (LSTM) units \citep{Hochreiter1997} on a language identification
task involving typologically and genetically different languages. The network was only given three features from the speech recordings to constrain the network to rely on speech rhythm: two amplitude envelopes -- one derived from the raw waveforms and one from a high-passed version of the waveforms -- and a variable identifying the voiced segments. 
Our assumption is that these features convey phonetic information poorly, but preserve the information related to rhythm, such as the alternation of weak and strong segments and the consonant-vowel segmentation. The network was trained on a large dataset of speech recordings in 21 languages collected from several open web databases. Our main question is whether the internal
representation of the trained model is consistent with the 
 rhythmic typology of languages.
To this end, we analyzed the network error structure, and we
used several visualization methods to build language maps from network output activations. We obtained successful language identification in 40\% of the cases, and the target language
was in the top 3 guesses in two-thirds of the cases. We also found that linear combinations of a sub-set of these activations presented strong correlation with established speech rhythm metrics, allowing interpretations in line with previous work.
The findings illustrate the strong potential of deep learning methods for speech rhythm 
research, providing a tool to strengthen the connection between hypotheses on the perception of language, and underlying statistical regularities present in the speech signal, even when these regularities are not easily quantified by more classical analytical models.

\section{Methods}\label{methods}

\subsection{Speech data}\label{speech-data}

The DNN model's task was to identify the language of 10-second speech recordings from 21 languages. To this end, a very large dataset was assembled by retrieving speech recordings from five open web databases. The main database
was the Common Voice dataset, an open source collaborative project
launched by Mozilla in 2017 collecting sentences read by web users
\citep{Ardila2020}, representing 66\% of the recordings. The other
databases were LibriVox (14.5\% of the recordings), a collaborative
project for the creation of public domain audiobooks; a set of radio
podcasts listed by the Wide Language Index for the Great Language Game
(WLI - 14.5\%) \citep{Skirgard2017}; VoxForge, a collaborative project
similar to the Common Voice initiative (4\%); and Tatoeba, another
collaborative database of read example sentences (1\%). The entire
dataset represents a large variety of speakers (25,000 in total) and
recording settings, a necessary condition for the neural network to
generalize to examples outside the training set. The data also includes different elicitation methods (reading of short sentences, reading of
running text, connected speech) although the reading of disconnected
sentences is the main elicitation method represented. The data was
 cleaned in several ways before being used to train the model. Files that excessively
contained music excerpts in the LibriVox audiobooks and the WLI podcasts
were removed. For Common Voice, the metadata was used to
exclude the recordings of the lowest quality, based on down- and up-votes as well as speech by non-native speakers, based on the `accent' field. We also chose not to include speakers with particular accents, especially French Canadian speakers
for French. However, we applied this process only to a limited extent, as we could not separate all regional varieties within a language due to lacking labels in the Common Voice metadata, e.g. European Portuguese and Brazilian Portuguese, even though they are known to exhibit different rhythmic characteristics. To create 10-second-long segments, files from the same speaker were concatenated together with a
0.3-sec pause between each file, if they were shorter than 10 seconds, whereas longer recordings (typically from
Librivox and the WLI) were split into 10-sec segments. We chose a relatively long segment duration to allow the neural network
to adapt to recording settings and to accumulate evidence over
several sentences. The resulting dataset contained 440,000 10-sec
recording segments in total, representing 1,200 hours of speech data. The
recording segments, however, were not evenly distributed across languages (Table
\ref{tab:datainfo}). The two most represented languages, German and
English, accounted for 30\% of the dataset. On the other end of the list,
Estonian, Japanese and Hungarian only have 3 000-4 000
segments (\textless{}1\%). This major bias in favor of the
most widespread languages was counterbalanced by adding weights to the
cost function during model training, as specified further on.

\subsection{Model inputs}\label{features}

\begin{figure}
\centering
\includegraphics[width=4.5in]{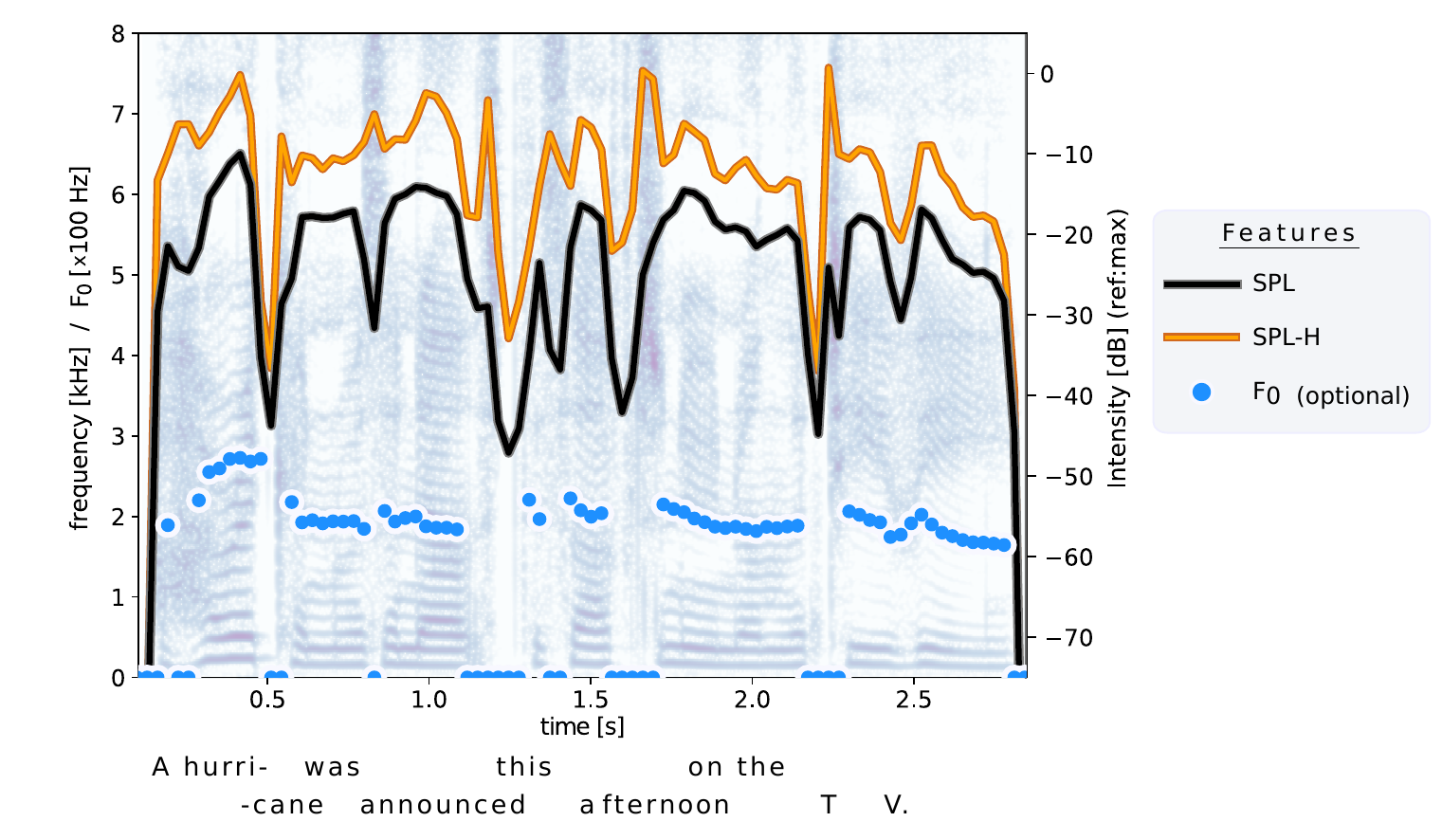}
\caption{\label{fig:example_data} Features for the language
identification task illustrated on the sentence: `a hurricane was
announced this afternoon on the TV' (from the Ramus corpus
\citep{ramus1999a}). SPL: Sound pressure level. SPL-H: Sound pressure level after the signal is passed through a gentle high-pass filter. F0 : the fundamental
frequency. In the main version of the model presented in the paper, only the voicing information is kept for the third dimension -- 1 for voiced segments ($F_0 \neq 0$); 0 for voiceless segments ($F_0=0$). The three features are
sampled at 31.25 Hz. The spectrogram of the sentence is shown
for guidance (background image).}
\end{figure}

Inputs extracted from the speech recordings were intentionally limited to force the model to rely primarily on prosodic cues. The inputs consisted of a three-dimensional feature vector presented at a sampling rate of 31.25 Hz. This sampling rate corresponds to time windows of 512 samples at 16 kHz, or 32 ms. The first two features were the average sound pressure level (SPL)
and the SPL level with a pre-emphasis on high frequencies
(SPL-H), both in dB, computed on the 32-ms time windows. The third feature was a variable indicating whether the windows belonged to voiced segments (0 for
voiceless segments, 1 otherwise). They are illustrated on a example in Fig.~\ref{fig:example_data}.

We chose these features on the basis of acoustic studies of prominence in Swedish \citep{Fant2000}. Specifically, the amplitude envelopes SPL and SPL-H delineate the main prosodic
boundaries and syllable contours. The local alternations between low and high intensity
segments provide additional clues about consonant-vowel (CV) transitions.
These transitions are further indicated by the addition of SPL-H, since most vowels
and consonants have a different balance between low and high
frequencies; for example, most fricatives primarily carry energy in the higher frequency range. Variations in intensity can also be an indication
of syllable prominence. The
difference between SPL-H and SPL also provides information about  spectral tilt,
which in turn is a correlate of vowel openness, vowel quality and stress
\citep{Sluijter1996, Fant2000}. The third feature naturally marks the transitions between voiced and unvoiced segments. The sampling frequency (31.25 Hz) was chosen so that the corresponding Nyquist frequency was of the same order as the rate of 10-15 phonemes per second in normal speech. 

Prior to the computation of the intensity measures (SPL and SPL-H), the signal amplitude of each recording was normalized to ensure that the dynamic ranges of all recordings were comparable. SPL-H is the intensity of the signal
after passing through a gentle high-pass filter with the following frequency response:

\[H(j\omega) = \frac{1+j \omega / \omega_A}{1+j \omega / \omega_B} \ ,\]

or, after applying the bilinear transform:

\[H(z) = \frac{(1+A)+(1-A)\, z^{-1}}{(1+B) + (1-B)\, z^{-1}} \ ,\]

with \(A=(2/\omega_A) f_s\), \(B=(2/\omega_B) f_s\),
\(\omega_A= 2\pi \times \ 200\) Hz, \(\omega_B= 2\pi \times \ 5000\) Hz,
and \(f_s\) the sampling frequency (16 kHz). This filter applies an
extra gain of around +20dB for frequencies above 1 kHz (see figure 12 in \citep{Fant2000} considering the same filter). To compute the third feature (voicing information), we used the Parselmouth library based on Praat to estimate the time course of the fundamental frequency F0 \citep{Jadoul2018, Boersma2018}. As a result, our code provides the option to include F0 as the third feature (Fig.~\ref{fig:example_data}), however we only kept the voicing information in the main version of the model presented in the Results section; i.e., voiceless segments are coded as 0, other segments are set to 1 irrespective of the F0 value. 
Training the
neural network was more difficult when F0 was included instead of the voicing information, especially for
the models with a small number of units that were in general preferred (see next subsection). Before being processed by the DNN, the features were normalized by applying a constant affine
transformation to each dimension so that each input was in the range [0,1].


\begin{table}[ht]
\centering
\begin{tabular}{llllll}
\toprule
Language   & Nb files        & Language    & Nb files       & Language   & Nb files      \\
\midrule
German     & 72,368 (16.3\%) & Russian     & 14,880 (3.3\%) & Polish \( \ast \)    & 6,832 (1.5\%) \\
English \( \ast \)   & 64,512 (14.5\%) & Mandarin    & 14,000 (3.1\%) & Danish     & 6,048 (1.4\%) \\
French \( \ast \)    & 54,816 (12.3\%) & Portuguese  & 14,112 (3.2\%) & Arabic     & 5,520 (1.2\%) \\
Catalan \( \ast \)   & 46,576 (10.5\%) & Korean      & 10,896 (2.4\%) & Turkish    & 5,376 (1.2\%) \\
Spanish \( \ast \)   & 43,584 (9.8\%)  & Swedish     & 10,320 (2.3\%) & Estonian   & 3,520 (0.8\%) \\
Italian \( \ast \)   & 30,416 (6.8\%)  & Dutch  \( \ast \)     & 10,640 (2.4\%) & Japanese \( \ast \)  & 3,856 (0.9\%) \\
Basque     & 15,600 (3.5\%)  & Finnish     & 8,320 (1.9\%)  & Hungarian  & 3,024 (0.7\%) \\
\bottomrule
\end{tabular}
\caption{\label{tab:datainfo} Number of 10-second speech recordings by language (total: 445,216) and corresponding percentage of the total dataset. Detailed statistics on the training and test datasets are provided in Supplementary Information (Tables Supp. 1 to Supp. 3). \( \ast \) = Language present in the Ramus corpus \citep{ramus1999a}.}
\end{table}

\subsection{Model architecture}\label{model-architecture}

\begin{figure}
\centering
\includegraphics[width=5.5in]{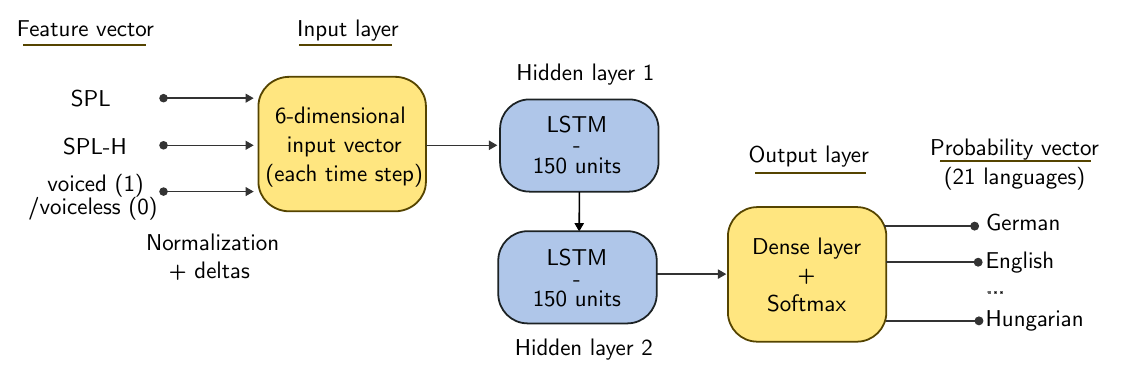}
\caption{\label{fig:diagram} Block diagram of the recurrent neural
network used for the language identification task. The input layer consists of the
three features and the associated deltas (differences between two time steps) sampled at 31.25 Hz. The two
hidden layers contain 150 LSTM units each; the model output after applying the softmax function can be interpreted as a probability vector of
the most likely languages identified by the network. }
\end{figure}

A simplified diagram of the main version of the recurrent neural network
implemented for the language identification task is shown in Fig.~\ref{fig:diagram}. The network inputs are vectors of dimension 6
regrouping the three normalized features presented in the previous subsection (values between 0 and 1) as well as
the associated deltas (differences between two time steps). The network
contains two hidden layers of 150 Long Short-term Memory (LSTM) units
with forget gates \citep{Gers2000}. The main benefit of LSTM networks over standard
recurrent neural networks is to avoid the vanishing gradient problem during training \citep{Hochreiter1997}. Each unit is associated with a hidden
state but also a cell state which serves as a memory. The action of the output layer consists in linearly combining the hidden states of the second LSTM layer and applying the softmax operation.
The result is an output vector \(\mathbf{\hat{y}}\) that can be interpreted as a probability distribution indicating the likelihood of each language in the dataset being the language of the current input. 
Considering that the languages are indexed from 1 to 21 (1= German, 2=English, etc.), the desired output for a single recording is represented by a one-hot vector \( \mathbf{y} \) that has only zeros except for the coordinate corresponding to the correct language;
i.e, of the form \(\mathbf{y}=\left[0, \cdots, 0, 1, 0, \cdots 0 \right]\). 
The recurrent neural network is then trained to minimize the cross-entropy between the desired and generated vectors \(H(\mathbf{y}, \mathbf{\hat{y}}) = -\sum_{i} y_i \log(\hat{y}_i)\), averaged over all time-steps. 

The model was
implemented using Keras and TensorFlow \citep{Abadi2016}. It was optimized by
batch gradient descent using the truncated backpropagation algorithm
with truncated sequences of 32 time steps corresponding to about one second. The 10-second recordings were therefore divided into 10 slices to compute the gradients, but the LSTM states were reinitialized only at the presentation of a new recording. 
Nesterov's accelerated gradient scheme was applied with a momentum of 0.9. The learning rate was initially set to 0.2 and was
reduced to 93\% of its value at each epoch. We used dropout
\citep{Srivastava2014} as a regularization method (recurrent dropout \citep{Semeniuta2016}:
0.1. dropout between recurrent layers: 0.1, dropout for the dense layer:
0.2). A data augmentation scheme was used by applying a gentle distortion to the inputs: for each new batch, the identity mapping {[}0,1{]} was replaced by the sum of six sigmoids evenly placed on the 
segment {[}0,1{]} whose maximum slope varied from the unity by a random
angle (standard deviation: \(\pi/10\); contraction factor for
the sigmoids: 15).

The dataset was split into a training and a test set (8\% of the data) with
different speakers to evaluate the performance of the model. The
classification error and top-3 error were monitored during training using TensorBoard \citep{Abadi2016}. The top-3 error is
the proportion of recordings for which the three most likely languages identified by the network do not include the correct label. The classification error and top-3 errors were only based on the model output at the end of each 10-second recording, not the intermediate outputs. 
The model was trained on 25 epochs; i.e., each recording was used 25 times during training to update the model. This number of epochs corresponded to the point where the accuracy on the test set stopped improving (early stopping) to limit overfitting. 
Since the distribution of recordings across languages and
speakers was not even for the training set, the overall contribution of each language and
speaker was set to match saturating functions. More explicitly, for a
given language, a variable \(n_{spk}\) was defined as
\(n_{spk}= \sum_{s} n(s)/(K1 + n(s))\) where \(n(s)\) is the number
of recordings for the speaker \(s\). Then each sample associated with
the speaker \(s\) was attributed the following weight in the loss
function: \(1/(K2 + n_{spk}) \cdot 1/(K1 + n(s))\), with \(K1=20\),
\(K2=5\).  The number of recordings by languages was more even for the test set (see Supplementary Figure Supp. 5).

It is important to note that performance was not the only criterion
for selecting the architecture and the hyperparameters. In particular, we oriented the choice of the architecture towards smaller
networks to force the model to learn shared representations across
languages. Although larger networks may exhibit better performance, they tend to produce more independent clusters as shown by visualization methods such as t-SNE
(see next subsection). This observation suggests that larger networks developed specialized parts for recognizing specific languages, i.e. they overfit the data, while rhythmic features should remain as general as possible. 
For the same reason, we combined different strategies of dropout to encourage the network to rely on a reduced number of activated units. 
Roughly, networks containing less than 100 units per layer had lower and more variable accuracy, whereas networks with over 200 units showed signs of overfitting to the languages of the dataset. 
We opted for 150 units per layer for the main version of the model as a trade-off between these two tendencies. Architectures with 3 hidden layers achieved similar
performance as the 2-layer model, but were more difficult to train, therefore we only consider 2-layer models in the Results section. 

\subsection{Visualization methods}\label{visualization-methods}

To inspect the model internal representation of languages after training, we carried out several visualization methods and clustering analyses based on the output probability vectors, using the following techniques:

\begin{itemize}
\item
  metric multidimensional scaling (MDS)
\item
  hierarchical clustering
\item
  t-distributed stochastic neighbor embedding (t-SNE)
  \citep{VanderMaaten2008}.
\end{itemize}

We used the implementations of scikit-learn for the three methods \citep{scikit-learn}. The
first two techniques require single high-dimensional vectors representing
each class. For those, we used histograms of the last layer activations computed on a
small balanced set of 6 720 recordings (320 recordings x 21 languages) sampled
randomly from the larger dataset. More explicitly, let \(z_{n}\) be
the output probability vector obtained at the end of the n-th recording; we
define the vector \(\omega_{i}\) representing the \(i\)-th
language such as
\([\omega_{i}]_n = \frac{ [z_{n}]_i}{\sum_k{[z_{k}]_i} }\), where
\([z_{n}]_i\) denotes the i-th component of the vector \(z_n\).
Several dissimilarity measures can be derived from the vectors
\(\omega\). The results presented in the paper were obtained by applying the square root to each component of the vectors (normalization of the Euclidian norm) and
by comparing them using the Euclidean distance (Hellinger distance) or
minus the logarithm of their cross-product (Bhattacharyya distance). The
pairwise comparisons were stored in a dissimilarity matrix that was then used as
input for the MDS or hierarchical clustering methods. The Bhattacharyya distance quantifies the overlap between two probability distributions. For instance, if two normal distributions with the same variance are separated by 2 (or 4) standard deviations (measured by the gap between their means), the corresponding Bhattacharyya distance is 0.5 (or 2).

The t-SNE method differs from MDS as it seeks to faithfully render
short- and medium-range proximities between points rather that to
capture the general topology, which is fundamentally different for
objects living in a high dimensional space. The t-SNE algorithm is based
on a probabilistic interpretation of proximity by assuming a random
sampling of point pairs for which nearby points are drawn with a higher
probability. Two probability distributions are therefore defined, one in
the initial space and the other in the low-dimensional embedding 2-d
space. The algorithm aims to minimize a cross-entropy term
between the two distributions. We refer the interested reader to \citet{Wattenberg2016} for more insights into the
t-SNE method and its interpretation. For this analysis, 150 recordings
by language were selected and the output probability vectors \(z_{n}\) (normalized through square root transformation) were used as the points to visualize. 
The t-SNE
algorithm has a free parameter, the perplexity, which corresponds to the
approximate number of neighbors of each point. The perplexity was set to
the number of points by class (150).

It is worth noting that we chose to keep the dropout regularization to perform the analyses described above, even though the dropout of network units is usually removed post-training. This choice was motivated by the known fact that DNNs tend to show overconfidence in their predictions even though the
classification accuracy does not warrant it. In the absence of dropout, the
correlation between vectors representing different languages would typically be low, hindering the visualization methods. A mathematical interpretation of this strategy is presented in \citet{Gal2016}.

\subsection{Comparison with speech
metrics}\label{comparison-with-speech-metrics}

The analyses mentioned in the previous paragraphs do not reveal which features the network relies on to make its predictions. The resulting visualizations also lack interpretable axes.
Previous work on rhythm metrics allowed for more direct explanations because the construction of the metrics was explicit, based on consonant-vowel segmentation. 
To fill the interpretative gap between the two approaches, we tested whether the features learned by the recurrent neural network were correlated with known rhythm metrics. 
To this end, we used the Ramus corpus \citep{ramus1999a}, in which sentences were already segmented into consonants (C) and vowels (V). The rhythm metrics considered were the following:

\begin{itemize}
\item
  \(\%V\): the proportion of time covered by the vocalic intervals,
  \(\Delta C\): the standard deviation of consonantal intervals,
  \(\Delta V\): the standard deviation of vocalic intervals
  \citep{ramus1999a}
\item
  Varcos: the same metrics as above but \(\Delta C\) (resp. \(\Delta V\)) is
  divided by the mean consonantal (resp. vocalic) interval duration
  \citep{Dellwo2006}
\item
  the \emph{pairwise variability indexes} (PVIs) \citep{Grabe2002}
\end{itemize}

Raw PVIs and normalized PVIs are defined by the following
equations: \[rPVI= \sum_{k=1}^{m-1} |d_{k+1} - d_{k}|/(m-1)\]

\[nPVI= \sum_{k=1}^{m-1} |\frac{d_{k+1} - d_{k}}{ (d_{k+1} + d_{k})/2}|/(m-1),\]

where \(d_k\) denotes the duration of the k-th consonantal or vocalic interval. Consistent with
the rest of the literature, we considered raw PVIs for consonantal
intervals (\(rPVI_C\)) and normalized PVIs for vocalic intervals
(\(nPVI_V\)).

The different metrics were computed for each sentence of the Ramus corpus for which the segmentation was available (160 sentences from the following 8 languages: French, English, Dutch, Polish,
Spanish, Italian, Catalan, Japanese). The same sentences were provided
as inputs to the neural network and the activations of the first and
second hidden layers at the end of the sentences were stored.

We employed two strategies to examine the correlations between the rhythm metrics and the model
activations. 
The first strategy was to test if the correlation between single cell activations and the rhythm metrics was significant using the Pearson coefficient (r) and the Bonferroni correction for multiple testing, assuming the normality of both activation and metric values. 
The second strategy consisted in mapping the rhythm metrics to linear combinations of cell
activations within a hidden layer. For this analysis, we used a linear regression model with a regularized least squares method due to limited available data. More specifically, we adopted the ElasticNet \citep{Zou2005}
implementation of scikit-learn \citep{scikit-learn}. 
The ElasticNet's objective function is of the form
\(1 /(2n) ||y - Xw||^2_2 + \alpha r_{1} ||w||_1 + \alpha (1 - r_{1}) ||w||^2_2/2\).
\(r_{1}\) was set to 0.2 and \(\alpha\) was determined through a cross-validation procedure 
with 7 folds. The metrics were normalized beforehand to ensure that the vectors $y$ had a quadratic norm of 1. The results of the regression analysis were then used to generate language maps embedded in 2-d feature spaces.

\section{Results}\label{results}

\subsection{Language discrimination}\label{lgid}

\begin{figure}
\centering
\includegraphics[width=\linewidth]{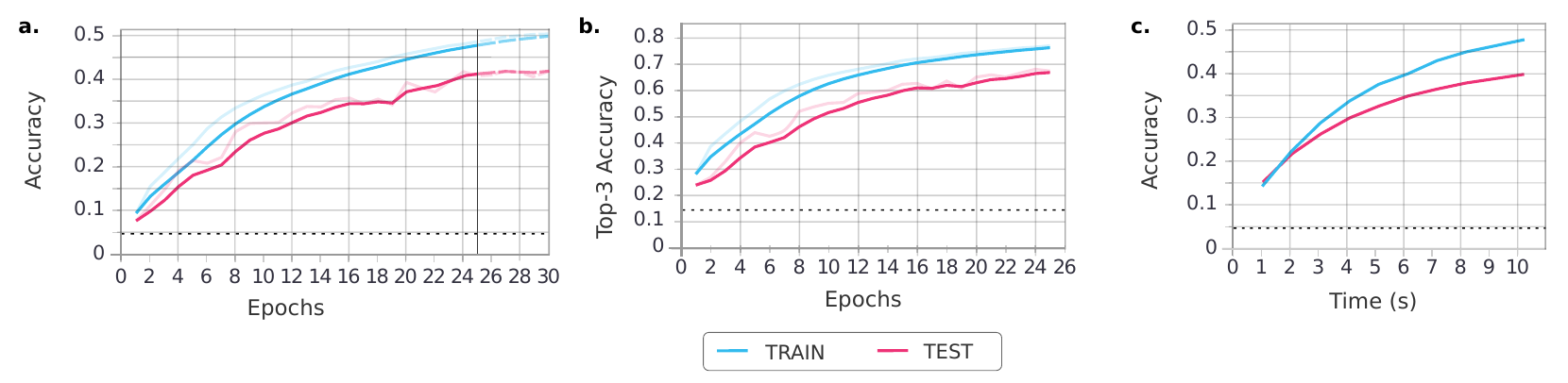}
\caption{\label{fig:accuracy} \textbf{(a)} Model accuracy and \textbf{(b)} top-3 accuracy during training as a function of epoch for the
language identification task. After 25 epochs corresponding to the trained version presented in the paper (early stopping), the test accuracy stopped improving. \textbf{(c)} Mean accuracy after training at different time points within the recordings (0= start of recordings, 10
sec=end of recordings). The horizontal dashed lines indicate the chance levels assuming equiprobability of the language classes. Light lines: raw numbers; dark lines: smoothed data; colored dashed lines: data between epochs 26 and 30 (for reference).}
\end{figure}

The trained model with $2 \times 150$ units was able to recognize the language of the 10-second
recordings 41\% of the time on the test set (on training set: 50\%), and the correct language was in the top 3 guesses of the network
67\% of the time (on training set: 80\%). 
Figure~\ref{fig:accuracy}a and \ref{fig:accuracy}b shows the evolution of accuracy during training; Fig.~\ref{fig:accuracy}c shows the average accuracy at different time points within the recordings. 
Although the training and test datasets have different speakers, they originate from the same databases and share correlated characteristics. The presented results are therefore likely higher than the actual performance  measured on a independent speech dataset. Indeed, the accuracy on 3-second recordings of the Ramus corpus is 18\% (top 3 accuracy: 42\%), compared to 25\% at 3 seconds on our dataset (Fig.~\ref{fig:accuracy}c).

A larger network of $2 \times 180$ units and no dropout for the last layer achieved an accuracy of 57\%
on the test set, and a top-3 accuracy of 79\%. With the addition of F0, the same model reached 61\% in accuracy and 82\% for top-3
accuracy. We show the results obtained with this higher-performing model in Supplementary Information (Fig. Supp. 11 to Supp. 15). 
The main difference is that the individual languages are more distant from each other in the activation-based representational spaces, 
indicating a higher level of specialized processing for each language. The results, however, align with those of the small model that the rest of the results presented below are based on.

\begin{figure}
\centering
\includegraphics[width=3.80000in]{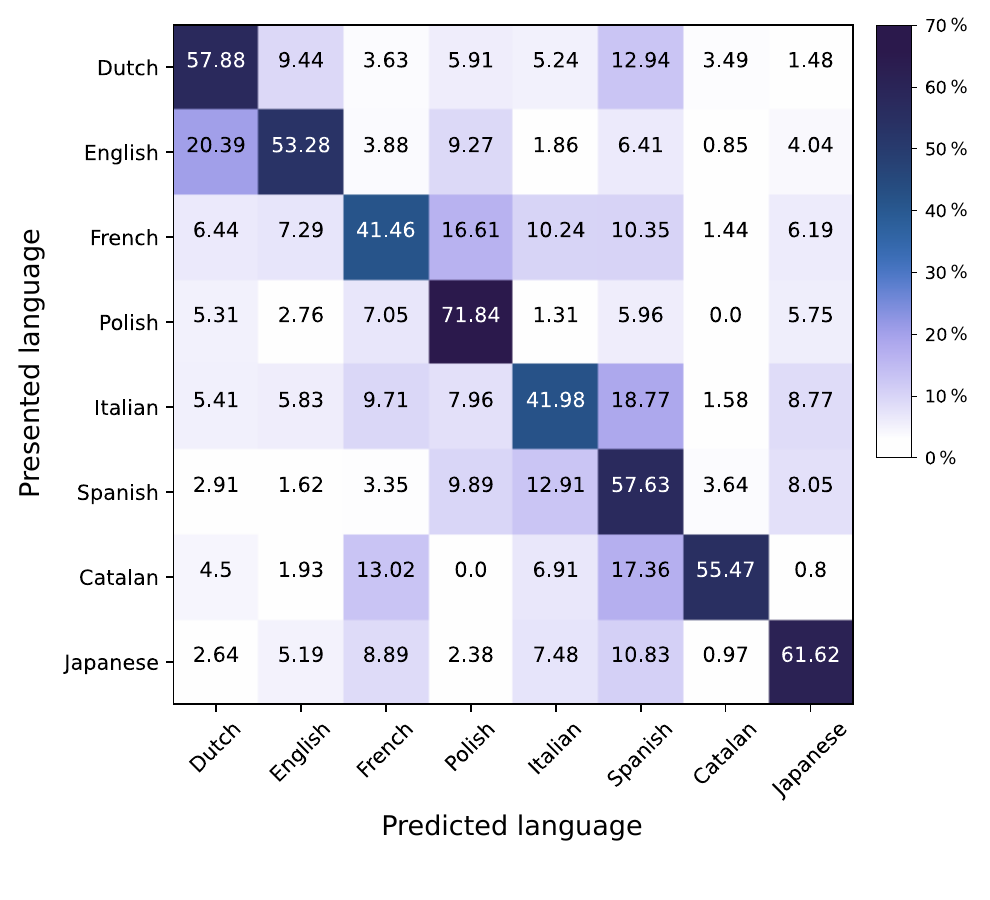}
\caption{\label{fig:confmatrix} Confusion matrix on the test set for a
limited number of languages (languages from the Ramus corpus
\citep{ramus1999a}). The figures correspond to percentages (normalized
by row). The entire confusion matrix is provided as Supplementary Fig. Supp. 6.}
\end{figure}

\begin{figure}
\centering
\includegraphics[width=\linewidth]{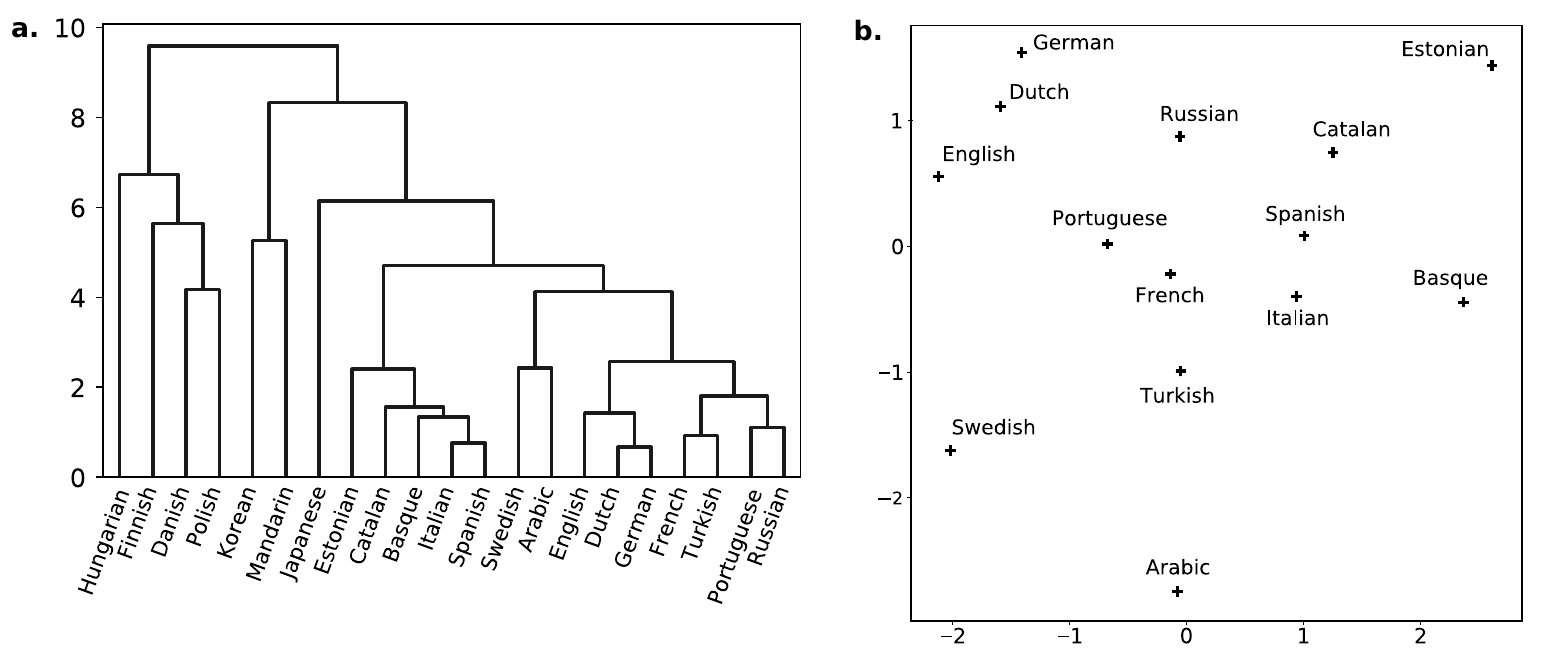}
\caption{\label{fig:bhat_figs} \textbf{(a)} Hierarchical clustering
dendrogram based on histograms of the DNN probability vector output using
the complete linkage method and the Bhattacharyya distance. \textbf{(b)}
Metric dimensional scaling (MDS) visualization for the languages in the
right branch of the dendrogram, also based on the Bhattacharyya distance
between activation histograms. MDS stress: 0.14.}
\end{figure}

The confusion matrix restricted to the eight
languages of the Ramus corpus is displayed in Fig.~\ref{fig:confmatrix}; the full matrix can be found as Supplementary Fig. Supp. 6. The confusion matrix alone does not reveal any clear cluster or groups of languages. To further investigate if such groups can be extracted from model confusion patterns, we carried out
proximity analyses on histograms computed from the activations of the
output layer. Fig.~\ref{fig:bhat_figs}a shows a dendrogram built from a
hierarchical clustering method with complete linkage using the
Bhattacharyya distance. The y-axis corresponds to the maximum distance between languages of two clusters when they
are merged to form a larger group. The dendrogram displayed represents only one possible version: the specific arrangement of languages can vary depending on the distance or the linkage method used, as well as the version of the neural network considered for analysis. 
This is especially true for the languages that are well discriminated by the neural network,
corresponding to the languages with a high junction point on the diagram
(including Asian languages). To illustrate this point, we show other versions of the dendrogram generated for the same DNN using an information-theoretic distance or a different linkage method in Supplementary Fig. Supp. 7.
The right part of the dendrogram illustrated in  Fig.~\ref{fig:bhat_figs}a, on the contrary, was found to be more robust to changes in method parameters. The first paired languages are Dutch and German, followed by Italian and Spanish (Bhattacharya distances: 0.71 and 0.76). The next pairs of languages are French and Turkish, then Portuguese and Russian (distances: 0.91 and 1.14). 

\begin{figure}
\centering
\includegraphics[width=4.20000in]{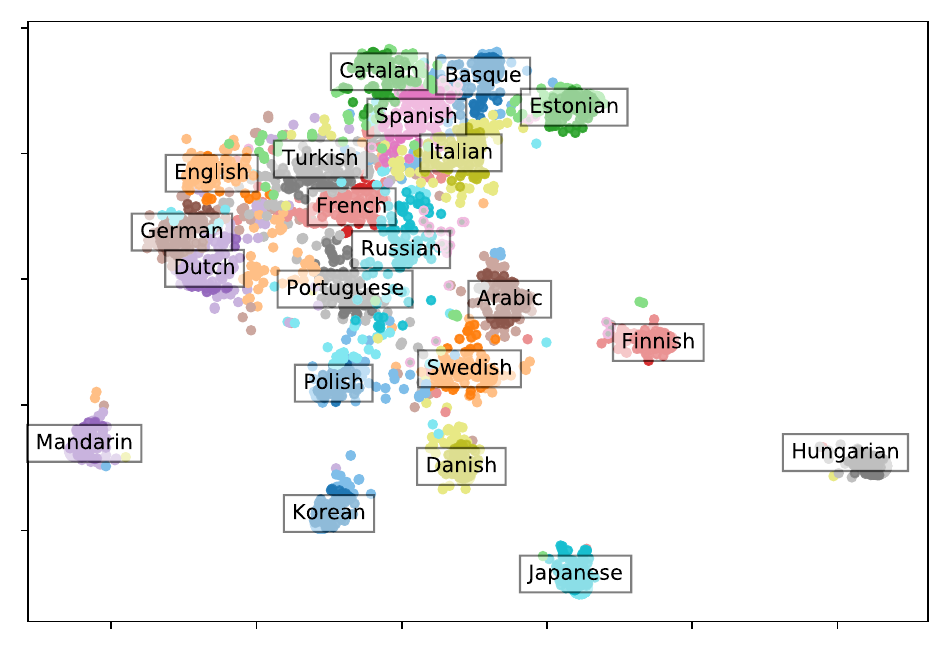}
\caption{\label{fig:tSNE} One output of the t-SNE algorithm based on the
output probability vectors of single samples, represented by each point,
and the Hellinger distance. Labels and colors correspond to the
predicted language, i.e.~the language as seen by the model. Number of
samples by classes and perplexity: 150.}
\end{figure}

The
representation of languages based on the network activations can be
further investigated using the MDS and t-SNE visualization tools (Fig.~\ref{fig:bhat_figs}b and Fig.~\ref{fig:tSNE}). The two figures provide similar results and clarify the relative positioning of the language groups previously mentioned. In particular, French, Turkish, Russian and Portuguese, are found to occupy an intermediate position between the stress-timed Germanic languages (English, German, Dutch), and the cluster dominated by syllable-timed languages (including Spanish and Italian). The MDS analysis in Fig.~\ref{fig:bhat_figs}b only included the languages corresponding to the right part of the dendrogram to achieve a stress (MDS cost function) below 0.2. By contrast, the t-SNE plot (Fig.~\ref{fig:tSNE}), which focuses on preserving local proximity rather than the global topology, includes all the languages of the dataset.
However, similar to the left part of the dendrogram, interpreting the results should be approached with caution for the languages that are well discriminated by the network, forming independent clusters in the t-SNE plot. Their relative positioning is  more sensitive to the samples provided as input or the algorithm's initialization compared to the denser central cluster.

\subsection{Correlations with rhythm metrics}\label{correlations}

To enhance the interpretability of our model, we tested whether 
some of the learned features, encapsulated by the hidden layer activations, were correlated with established rhythm metrics. On the Ramus corpus \citep{ramus1999a}, we found that
the rhythm metric presenting the highest correlation with a single cell
activation was \(\%V\) (r=0.52, p\textless{}1e-4 assuming independent
Gaussian variables with Bonferroni correction), followed by \(nPVI_V\)
(r=0.47, p\textless{}1e-4), the PVI index for vocalic intervals. The
fact that the highest correlation was found for \(\%V\) (proportion of
duration covered by vocalic intervals) was not surprising as it is a
relatively simple feature to learn. The other metrics in the order of
decreasing Pearson coefficient were: \(\Delta C\) (r=0.45,
p\textless{}1e-4), \(rPVI_C\) (r=0.40, p\textless{}1e-4), \(\Delta V\)
(r=0.37, p\textless{}1e-3), and the Varcos (r = 0.3, p=0.05). Most 
high correlations between rhythm metrics and cell activations were
found within the second hidden layer. Supplementary Information provides two language maps based on these single cell activations (Supplementary Fig. Supp. 8 and Supp. 9).

\begin{figure}
\centering
\includegraphics[width=\linewidth]{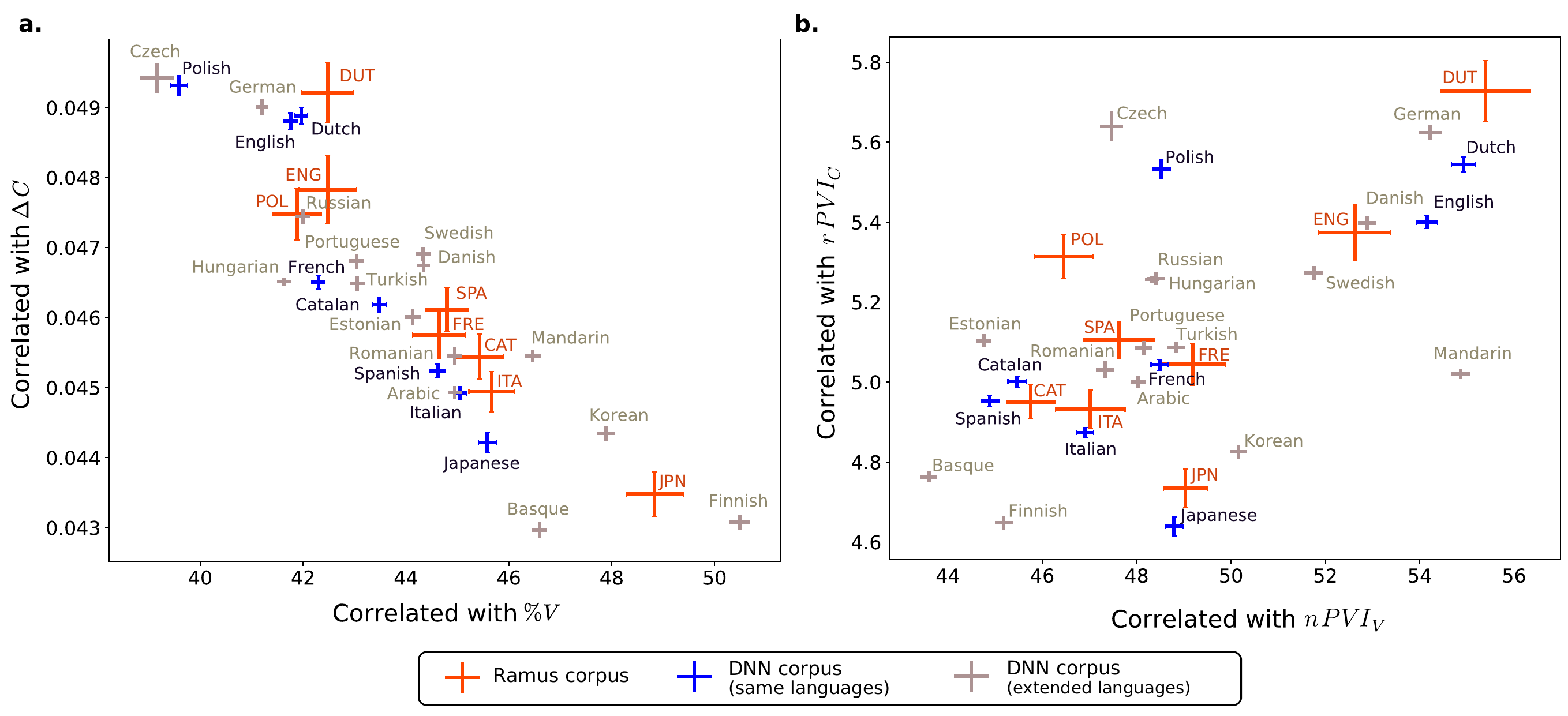}
\caption{\label{fig:map_metrics} Language maps in spaces defined by linear combinations of learned features (hidden layer activations) correlated with rhythm metrics
(based on the Ramus corpus and ElasticNet). The axes
are defined by the correlates of \textbf{(a)} (\(\%V\), \(\Delta C\)) and 
of  \textbf{(b)} (\(nPVI_V\), \(rPVI_C\)). The error bars correspond to standard
error, for data averaged over the Ramus corpus (red crosses,
short form labels) and over 7,000 recordings from our dataset 
(blue and grey crosses).}
\end{figure}

Because the information corresponding to rhythm metrics may be
distributed across several cells, we also considered linear combinations of neural units using the results of linear regressions between
the activations of the second hidden layer and the rhythm metrics. We
used this analysis to generate plots analogous to maps based on rhythm
metrics. The maps for the correlates of (\(\%V\), \(\Delta C\))
and (\(nPVI_V\), \(rPVI_C\)) are shown in Fig.~\ref{fig:map_metrics}. The figure for the correlates of (\(\Delta V\), \(\Delta C\)) is similar to the figure for the PVIs; it can be found as Supplementary Fig. Supp. 10 online. The red crosses represent the combination of activations computed on 
the Ramus corpus. Since these recordings were used to find the
correlates, the red crosses are naturally arranged in a similar manner to the maps based on rhythm metrics using the same corpus \citep{Ramus2002}. The
blue crosses represent the same languages but with scores computed on a
larger corpus (320 recordings by languages from our dataset). The 
overall proximity of red and blue crosses for the same languages gives us confidence that the regression model was still valid beyond the initial dataset. This can be quantified by comparing the variance between languages to the variance within each language, which also gives an indication of the variability of the captured features. In decreasing order of the variance ratio, we find the correlates of \(rPVI_C\) (F(7,8)=26, p<1e-4), \(nPVI_V\) (F(7,8)=25, p<1e-4), \(\Delta C\) (F(7, 8)=16, p<1e-3), and \( \% V\) (F(7,8)=5.2, p=0.02). After this step, all the other languages of the DNN
model were included to provide a map with the 21 languages (gray crosses). We also added Czech and Romanian that were not used for the training of the DNN due to insufficient data -- note that Czech was most often
classified as Polish (55\% of the times) by the DNN and Romanian as Russian (25\% of the times).

A natural question that can arise from the analyses with the correlated features is how much information used by the neural
network for the discrimination task is available in these feature spaces. To address this question, we conducted a quadratic discriminant
analysis (QDA) in the space composed by the correlates of the four
metrics (\(\%V\), \(\Delta C\), \(nPVI_V\), \(rPVI_C\)). The QDA
classifier achieved an accuracy of 36\% on the training set (top-3: 60\%) and 25\% on the test set (top-3: 47.5\%). These scores represent 60\% to 80\% of the performance achieved by the DNN. We also tested the QDA model restricted to the five languages included in Loukina et al. \citep{Loukina2011} (except for Greek replaced by Spanish). We found an accuracy of 50\% on the test set, comparable to the average score of 46\% reported by the authors for classifiers using four rhythm metrics on recordings of entire paragraph readings. On the test set with the same restriction of languages, the neural network achieved 62\% accuracy.


\section{Discussion}\label{discussion}

We developed a recurrent neural network to identify languages on the basis of rhythmic features. We could investigate how languages were related or separated by the network using clustering and visualization techniques based on the network output probability vectors. A third of the languages, corresponding to the left branch of the dendrogram in Fig.~\ref{fig:bhat_figs}, were individually well separated, and their arrangement relative to the other languages was variable. The clustering of the other part of the languages could be interpreted more consistently. The Germanic stress-timed languages (English, Dutch, German) were systematically grouped together, as were Spanish and Italian (prototypical syllable-timed languages). Considering larger clusters on the dendrogram, the results were overall consistent with the traditional dichotomy of languages, but presented some fine-grained differences. Spanish and Italian were grouped together with Basque, Catalan and Estonian, while the latter two languages have sometimes
been described as having a mixed rhythm \citep{Grabe2002}. The Germanic languages formed a larger group with other stress-timed languages (e.g., Portuguese, Russian), but this group also contained Turkish and French, generally considered as syllable-timed. The MDS or t-SNE visualizations however positioned French and Turkish, in addition to Portuguese and Russian, between the groups of stress- and syllable-timed languages.

The regression analysis between the hidden states of the recurrent neural network and rhythm metrics, first conducted on the languages of the Ramus corpus \citep{ramus1999a} and extended to the 21 languages of our dataset, revealed that several dimensions of the DNN inner representation encodes features highly correlated with rhythm metrics, allowing interpretations in line with previous work. Figure~\ref{fig:map_metrics} shows the placement of the languages on feature maps whose axes are correlated with these metrics. The correlates of (\(\%V\), \(\Delta C\)) (Fig.~\ref{fig:map_metrics}a) exhibit a stress-timed to
syllable-timed gradient, with stress-timed languages associated with a
lower value for the correlate of \(\%V\), the percentage of time covered 
by vocalic intervals, and a higher value for the correlate of
\(\Delta C\), supposedly reflecting syllable structure complexity. Most of the points within the plot are found on the same diagonal, consistent with the previous observation that \(\%V\) and \(\Delta C\) are highly correlated\citep{Ramus2002}, making it difficult to distinguish the properties represented by the two axes. Despite the apparent gradient of stress- and syllable-timing, it cannot be concluded that all languages are distributed according to this rule based on the traditional categorization of languages. For example, Czech appears at the upper left corner of the plot, but is sometimes classified as syllable-timed. The positioning of Czech in the plots of Fig.~\ref{fig:map_metrics} is however consistent with a previous study which found that Czech is within the cluster of stress-timed languages for \(\%V\) and \(rPVI_C\), but not for \(nPVI_V\) \citep{Dellwo2007}. Conversely, Arabic
is generally considered to be stress-timed but is found in the lower right
quadrant in panel~A. The points computed on the Ramus corpus (red crosses) exhibit the three separated classes that were described by the authors
of the original study (syllable-, stress- and mora-timed languages
\citep{Ramus2002}), but this separation in three classes does not
generalize to our dataset including more languages. The distribution of
points is rather in line with the idea of a continuum  advocated by Dauer, among others
,\citep{Dauer1983, Nespor1990} and supported by other studies based on rhythm
metrics \citep{Grabe2002, Mairano2011}. Note that, if we were to adhere to the separation in three classes suggested in Ramus et al., Finnish, Basque and Korean would be categorized as presenting mora-timed features. The figure with the correlates of the PVIs (Fig.~\ref{fig:map_metrics}b) exhibits a similar gradient
with syllable-time languages occupying the left lower part of the figure
and stress-timed languages occupying the right upper part. This
configuration is assumed to reflect the fact that stress-timed languages
are marked by greater variability in vocalic and inter-vocalic interval
lengths. However, for this second figure, a few languages depart from
the main diagonal. This is the case especially for Czech and Polish,
which occupy the upper left quadrant of the plot. This result aligns with previous studies using rhythm metrics \citep{Ramus2002, Grabe2002, Dellwo2007} and is consistent with the observation that these two languages are characterized by complex consonantal clusters but no vowel reduction at
normal speech rates.

While we have provided evidence that DNNs are useful tools to detect rhythmic features of languages, our work has certain limitations. We have used a strongly degraded signal, leaving only the amplitude envelope and voicing information, in order to restrict the input to prosodic features. We know, however, that even a severely degraded speech signal consisting of only a few amplitude envelopes can provide enough phonetic information for adults to comprehend speech under ideal listening conditions, i.e., in silence \citep{Shannon1995}. This has motivated our choice to use medium-sized neural networks instead of larger networks that would integrate more phonetic or linguistic knowledge. But, even without a fine-grained processing of phonetic features, the DNN could still capture broad phonological regularities. For example, Portuguese and Russian could be close in the representational space, not only because of prosody, but because of other similarities, such as the extensive use of sibilant fricatives, easily recognized as weak segments with spectral energy 
mainly contained in the medium or high frequencies. In that regard, the correlation analysis with rhythm metrics helped to disentangle contributing factors and provide more interpretation to the visualisations derived from the network activations. This does not completely resolve the issue -- in fact, even rhythm metrics have been criticized by a few authors for relying too much on phonological factors without a direct link to prosody \citep{Kohler2009, Arvaniti2009}. As we also argue in the next paragraphs, these valid points should not hold too strict standards for exploratory work in speech rhythm. Since the works of Dauer \citep{Dauer1983} and Bertinetto \citep{Bertinetto1989} who promoted the view that phonology and rhythm are not independent but intertwined dimensions of speech production and perception, the search for phonological or acoustic correlates of linguistic rhythm has been productive in finding measurable bases for the intuitive notion of rhythmic classes among languages. The present study fully embraces this line of research by relying on a statistical language-identification model with only acoustic features as its input. It is important to note that while this approach has several merits 
(e.g., avoiding reliance on a priori segmental division of speech, incorporating aspects of phone quality or co-articulation), models that incorporate more abstract elements of phonological segmentation may be equally valid to investigate speech rhythm. As an example, a DNN model based on the durations of phonetic segments and CV clusters was developed to identify the prosody of native vs non-native English speakers \citep{Kyriakopoulos2019}. The extension and analysis of such models could provide complementary and valuable information about the statistical regularities related to speech rhythm.

Our approach exploits the remarkable effectiveness of recurrent neural networks for pattern-recognition tasks when a large
number of labeled examples is available for training \citep{Beaufays2014, Karpathy2015}. Among the multiple strengths of DNNs, they are efficient in 
handling factors of intra-class variability, such as changes in speech rate, a known common issue for rhythm metrics \citep{Ramus2002, Arvaniti2009, Mairano2011}. By relying on a DNN instead of explicit phonological metrics, we could bypass the costly task of manually annotating consonant-vowel segments. That way, more languages could be included in the feature maps. But the value of DNNs extends well beyond these practical advantages. DNNs have the capacity to learn and generalize multiple complex patterns from statistical regularities found in the training data, achieving a greater degree of expressivity than metrics, which by contrast are aggregate measures coding for a coarse aspect of timing \citep{Arvaniti2009, Turk2013}. This allows our model to reach good accuracy at the level of a few sentences, while classifiers based on rhythm metrics typically work on larger sets or considering a reduced number of rhythmically distinct languages \citep{ramus1999a, Arvaniti2009}. We showed that a QDA classifier trained on the 4th-dimensional DNN feature subspace correlated with speech metrics (Fig.~\ref{fig:map_metrics}) matched the performance of comparable classifiers using rhythm metrics \citep{Loukina2011}, but on recordings of shorter durations. These low-dimensional classifiers were outperformed by the full DNN model (62\% accuracy vs 50\% accuracy considering five languages).

DNNs shift modeling from low-dimensional analytical approaches to statistical models that have superior predictive power but are also more opaque -- a criticism of DNNs being that they do not offer a straightforward interpretation of how they map 
 inputs and outputs (commonly referred to as the 'black-box' problem). For this reason, the use of DNNs in cognitive science represents a major paradigm shift for the modeling of perception or sensory systems 
\citep{Kell2019, Cichy2019, Storrs2019}. In this last paragraph, we mention several important aspects of this methodological evolution that provide further context for our approach. First, although the lack of interpretability is a legitimate criticism of DNNs, they are also well suited for post-hoc analyses that mitigate this issue \citep{Cichy2019, Storrs2019}. The linear mapping of features with rhythm metrics in this study is one example, but some DNNs can be analyzed in much greater detail, as it has been done in visual object recognition\citep{Cammarata2020}. Much information about DNN behavior can be gained from running the model on custom-made stimuli:  in our case, these could be stereotypical rhythmic patterns or resynthesized speech stimuli \citep{ramus1999c}. This type of approach has been fruitful for investigating the sensitivity of DNNs to shape and how it compares to visual perception \citep{Kubilius2016}. A second point is that some caution is required when interpreting the results of DNN models, especially when considering them as possible models of perception. Although a coarse measure of model performance can be reminiscent of error patterns found in perception studies (e.g., the discrimination abilities of young infants in our case \citep{ramus1999a}), DNNs present biases that human subjects do not have. This is now well-documented, again in the area of visual perception \citep{Wichmann2023}.
Another cautionary note that certainly applies to our study is that appealing
visualizations generated during DNN analyses are not sufficient for building scientific knowledge \citep{Leavitt2020}. They can give a sense of intuitive knowledge about the DNN behavior but they lack the rigor of more formal approaches. 
This view has to be nuanced by recognizing that the analysis of DNN behavior provides rich information about the statistical structure of complex stimuli that can lead to new developments and hypotheses on related cognitive concepts \citep{Kell2019}. Cichy and Kaiser reinforced this point by arguing that, although the drawbacks of DNNs complicate their use as explanatory models, they are invaluable tools for exploring cognitive theories that are not yet fully established \citep{Cichy2019}. 

We think that Cichy and Kaiser's argument is particularly relevant to speech rhythm research: despite a long history of developments, the field is still in an exploratory phase, due to the complexity of the underlying phenomena and the lack of a powerful tool to connect the intuition of linguists with concrete 
statistical regularities in the speech signal. Research on speech rhythm is at a stage where it would certainly benefit from creative and diverse approaches using DNN models, that will have to be followed by more formal verifications of hypotheses and efforts to converge to a much needed\citep{Turk2013} unified view of speech rhythm. As in other areas of cognitive science, these advances would be facilitated by setting up interdisciplinary projects that bring together machine-learning experts and psychologists (and/or linguists), and by fostering the sharing of resources and models across labs \citep{Storrs2019}.

\section*{Code and data availability}

The code for training and running the model is available on \href{https://zenodo.org/doi/10.5281/zenodo.10211058}{https://zenodo.org/doi/10.5281/zenodo.10211058}. We also provide the model weights for the main version discussed in the paper, as well as the regression coefficients for the comparison with speech metrics. 

The training data was retrieved from publicly available datasets ( 
\href{https://commonvoice.mozilla.org/en/datasets}{https://commonvoice.mozilla.org/en/datasets}, \\ \href{https://github.com/larsyencken/wide-language-index}{https://github.com/larsyencken/wide-language-index},  
\href{https://www.voxforge.org/}{https://www.voxforge.org/},
\href{https://librivox.org/}{https://librivox.org/},
\href{https://tatoeba.org/}{https://tatoeba.org/}). The data from Ramus et al. 1999 was obtained from the authors of that study upon request. The parameters for the main version of the model discussed in the paper and the linear regression coefficients for the comparison with speech metrics are provided with the code.

\section*{Acknowledgements}

We wish to thank Jean-Pierre Nadal for his valuable support as a member of the ``SpeechCode'' project and as F.D's supervisor during his PhD. We gratefully acknowledge the support of Violette Daures in carrying out a part of the analyses on the Ramus corpus. We also thank Christian Lorenzi and Ramon Guevara, two other members of the ``SpeechCode'' project, for their insightful input. Thanks to Franck Ramus for providing
us with the speech samples and labeled data. This work
was partly funded by the project ``SpeechCode'' of the French National
Research Agency, the ANR, contract ANR-15-CE37-0009-03
(\href{http://www.agence-nationale-recherche.fr/}{http://www.agence-nationale-recherche.fr}), as well as the ECOS- Sud action nr. C20S02, the ERC Consolidator Grant 773202 ``BabyRhythm'', the ANR’s French Investissements d’Avenir – Labex EFL Program under Grant [ANR-10-LABX-0083], the Italian Ministry for Universities and Research FARE grant nr. R204MPRHKE, the European Union Next Generation EU NRRP M6C2 - Investment 2.1 - SYNPHONIA Project, as well as the Italian Ministry for Universities and Research PRIN grant nr. 2022WX3FM5 to JG. A part of this work was conducted while F.D. was on a
postdoctoral fellowship supported by Fondation Pour l'Audition (FPA
RD-2019-3).

\bibliographystyle{unsrtnat}


\begin{thebibliography}{60}
\providecommand{\natexlab}[1]{#1}
\providecommand{\url}[1]{\texttt{#1}}
\expandafter\ifx\csname urlstyle\endcsname\relax
  \providecommand{\doi}[1]{doi: #1}\else
  \providecommand{\doi}{doi: \begingroup \urlstyle{rm}\Url}\fi

\bibitem[Nazzi et~al.(1998)Nazzi, Bertoncini, and Mehler]{Nazzi1998}
T~Nazzi, J~Bertoncini, and Jacques Mehler.
\newblock Language discrimination by newborns: Toward an understanding of the
  role of rhythm.
\newblock \emph{Journal of experimental psychology. Human perception and
  performance}, 24\penalty0 (3):\penalty0 756--66, June 1998.

\bibitem[Mehler et~al.(2000)Mehler, Christophe, and Ramus]{Mehler2000}
Jacques Mehler, Anne Christophe, and Franck Ramus.
\newblock What we know about the initial state for language.
\newblock \emph{Image, Language, Brain: Papers from the First Mind-Brain
  Articulation Project Symposium}, \penalty0 (33 1):\penalty0 51--75, 2000.

\bibitem[Cutler(1994)]{Cutler1994}
Anne Cutler.
\newblock Segmentation problems, rhythmic solutions.
\newblock \emph{Lingua}, 92:\penalty0 81--104, April 1994.
\newblock ISSN 0024-3841.
\newblock \doi{10.1016/0024-3841(94)90338-7}.

\bibitem[Pike(1945)]{Pike1945}
Kenneth~L. Pike.
\newblock \emph{The {{Intonation}} of {{American English}}}.
\newblock {University of Michigan (Ann Arbor)}, July 1945.
\newblock \doi{10.2307/409880}.

\bibitem[Abercrombrie(1967)]{Abercrombrie1967}
D~Abercrombrie.
\newblock \emph{Elements of {{General Phonetics}}}.
\newblock {University of Edinburgh}, 1967.

\bibitem[Roach(1982)]{Roach1982}
Peter Roach.
\newblock On the distinction between 'stress-timed' and 'syllable-timed'
  languages.
\newblock \emph{Linguistics Controversies}, pages 73--79, 1982.

\bibitem[Dauer(1983)]{Dauer1983}
Rebecca~M. Dauer.
\newblock Stress-timing and syllable-timing reanalyzed.
\newblock \emph{Journal of Phonetics}, 11:\penalty0 51--62, 1983.

\bibitem[Langus et~al.(2017)Langus, Mehler, and Nespor]{Langus2017}
Alan Langus, Jacques Mehler, and Marina Nespor.
\newblock Rhythm in language acquisition.
\newblock \emph{Neuroscience \& Biobehavioral Reviews}, 81:\penalty0 158--166,
  October 2017.
\newblock ISSN 0149-7634.
\newblock \doi{10.1016/j.neubiorev.2016.12.012}.

\bibitem[Terken and Hermes(2000)]{Terken2000}
Jacques Terken and Dik Hermes.
\newblock The {{Perception}} of {{Prosodic Prominence}}.
\newblock In \emph{Prosody: {{Theory}} and {{Experiment}}}, pages 89--127.
  2000.
\newblock \doi{10.1007/978-94-015-9413-4_5}.

\bibitem[Bertinetto(1989)]{Bertinetto1989}
Pier Bertinetto.
\newblock Reflections on the dichotomy `stress' vs.`syllable-timing'.
\newblock \emph{Revue de phon{\'e}tique appliqu{\'e}e}, 91\penalty0
  (93):\penalty0 99--130, 1989.

\bibitem[Kohler(2009{\natexlab{a}})]{Kohler2009a}
Klaus~J Kohler.
\newblock Rhythm in {{Speech}} and {{Language}}.
\newblock \emph{Phonetica}, 66\penalty0 (1-2):\penalty0 29--45,
  2009{\natexlab{a}}.
\newblock ISSN 0031-8388.
\newblock \doi{10.1159/000208929}.

\bibitem[Cumming and Nolan(2010)]{Cumming2010}
Ruth~Elizabeth Cumming and F.~Nolan.
\newblock \emph{Speech Rhythm: The Language-Specific Integration of Pitch and
  Duration}.
\newblock PhD thesis, 2010.

\bibitem[Turk and {Shattuck-Hufnagel}(2013)]{Turk2013}
Alice Turk and Stefanie {Shattuck-Hufnagel}.
\newblock What is speech rhythm? {{A}} commentary on {{Arvaniti}} and
  {{Rodriquez}}, {{Krivokapi{\'c}}}, and {{Goswami}} and {{Leong}}.
\newblock \emph{Laboratory Phonology}, 4\penalty0 (1):\penalty0 93--118, 2013.
\newblock ISSN 1868-6346.
\newblock \doi{10.1515/lp-2013-0005}.

\bibitem[Ramus et~al.(1999)Ramus, Nespor, and Mehler]{ramus1999a}
Franck Ramus, Marina Nespor, and Jacques Mehler.
\newblock Correlates of linguistic rhythm in the speech signal.
\newblock \emph{Cognition}, 73\penalty0 (3):\penalty0 265--292, 1999.
\newblock ISSN 00100277.
\newblock \doi{10.1016/S0010-0277(99)00058-X}.

\bibitem[Grabe and Low(2002)]{Grabe2002}
Esther Grabe and Ee~Ling Low.
\newblock Durational variability in speech and the {{Rhythm Class Hypothesis}}.
\newblock In \emph{Laboratory {{Phonology}} 7}, pages 515--546. 2002.
\newblock ISBN 978-3-11-019710-5.
\newblock \doi{10.1515/9783110197105}.

\bibitem[Moon et~al.(1993)Moon, Cooper, and Fifer]{Moon1993}
Christine Moon, Robin~Panneton Cooper, and William~P. Fifer.
\newblock Two-day-olds prefer their native language.
\newblock \emph{Infant Behavior and Development}, 16\penalty0 (4):\penalty0
  495--500, October 1993.
\newblock ISSN 0163-6383.
\newblock \doi{10.1016/0163-6383(93)80007-U}.

\bibitem[Gervain and Mehler(2010)]{gervain10}
Judit Gervain and Jacques Mehler.
\newblock Speech perception and language acquisition in the first year of life.
\newblock \emph{Annual review of psychology}, 61:\penalty0 191--218, 2010.

\bibitem[Kyriakopoulos et~al.(2019)Kyriakopoulos, Knill, and
  Gales]{Kyriakopoulos2019}
Konstantinos Kyriakopoulos, Kate~M. Knill, and Mark~J.F. Gales.
\newblock A {{Deep Learning Approach}} to {{Automatic Characterisation}} of
  {{Rhythm}} in {{Non-Native English Speech}}.
\newblock In \emph{Interspeech 2019}, pages 1836--1840. {ISCA}, September 2019.
\newblock \doi{10.21437/Interspeech.2019-3186}.

\bibitem[Arvaniti(2009)]{Arvaniti2009}
Amalia Arvaniti.
\newblock Rhythm, {{Timing}} and the {{Timing}} of {{Rhythm}}.
\newblock \emph{Phonetica}, 66:\penalty0 46--63, 2009.
\newblock \doi{10.1159/000208930}.

\bibitem[Wiget et~al.(2010)Wiget, White, Schuppler, Grenon, Rauch, and
  Mattys]{Wiget2010}
Lukas Wiget, Laurence White, Barbara Schuppler, Izabelle Grenon, Olesya Rauch,
  and Sven~L. Mattys.
\newblock How stable are acoustic metrics of contrastive speech rhythm?
\newblock \emph{The Journal of the Acoustical Society of America}, 127\penalty0
  (3):\penalty0 1559--1569, March 2010.
\newblock ISSN 0001-4966.
\newblock \doi{10.1121/1.3293004}.

\bibitem[Arvaniti(2012)]{Arvaniti2012}
Amalia Arvaniti.
\newblock The usefulness of metrics in the quantification of speech rhythm.
\newblock \emph{Journal of Phonetics}, 40\penalty0 (3):\penalty0 351--373, May
  2012.
\newblock ISSN 00954470.
\newblock \doi{10.1016/j.wocn.2012.02.003}.

\bibitem[Rathcke and Smith(2015)]{Rathcke2015}
Tamara~V. Rathcke and Rachel~H. Smith.
\newblock Speech timing and linguistic rhythm: {{On}} the acoustic bases of
  rhythm typologies.
\newblock \emph{The Journal of the Acoustical Society of America}, 137\penalty0
  (5):\penalty0 2834--2845, May 2015.
\newblock ISSN 0001-4966.
\newblock \doi{10.1121/1.4919322}.

\bibitem[Yamins and DiCarlo(2016)]{Yamins2016}
Daniel L~K Yamins and James~J DiCarlo.
\newblock Using goal-driven deep learning models to understand sensory cortex.
\newblock \emph{Nature Neuroscience 2016 19:3}, 19\penalty0 (3):\penalty0
  356--365, February 2016.
\newblock ISSN 1546-1726.
\newblock \doi{10.1038/nn.4244}.

\bibitem[Kell and McDermott(2019)]{Kell2019}
Alexander~Je Kell and Josh~H McDermott.
\newblock Deep neural network models of sensory systems: Windows onto the role
  of task constraints.
\newblock \emph{Current Opinion in Neurobiology}, 55:\penalty0 121--132, 2019.
\newblock \doi{10.1016/j.conb.2019.02.003}.

\bibitem[Storrs and Kriegeskorte(2019)]{Storrs2019}
Katherine~R Storrs and Nikolaus Kriegeskorte.
\newblock Deep {{Learning}} for {{Cognitive Neuroscience}}.
\newblock In M~Gazzaniga, editor, \emph{The {{Cognitive Neurosciences}}, 6th
  {{Edition}}}. {MIT Press}, 2019.
\newblock ISBN 1903.01458v1.

\bibitem[LeCun et~al.(2015)LeCun, Bengio, and Hinton]{LeCun2015}
Yann LeCun, Yoshua Bengio, and Geoffrey Hinton.
\newblock Deep learning.
\newblock \emph{Nature}, 521\penalty0 (7553):\penalty0 436--444, May 2015.
\newblock ISSN 0028-0836, 1476-4687.
\newblock \doi{10.1038/nature14539}.

\bibitem[Krizhevsky et~al.(2017)Krizhevsky, Sutskever, and
  Hinton]{Krizhevsky2017}
Alex Krizhevsky, Ilya Sutskever, and Geoffrey~E Hinton.
\newblock {{ImageNet}} classification with deep convolutional neural networks.
\newblock \emph{Communications of the ACM}, 60\penalty0 (6):\penalty0 84--90,
  2017.
\newblock ISSN 15577317.
\newblock \doi{10.1145/3065386}.

\bibitem[Yu and Deng(2015)]{Yu2015}
Dong Yu and Li~Deng.
\newblock \emph{Automatic {{Speech Recognition}} : {{A Deep Learning
  Approach}}}.
\newblock {Springer London}, {London}, 2015.
\newblock \doi{10.1007/978-1-4471-5779-3}.

\bibitem[Hochreiter and Schmidhuber(1997)]{Hochreiter1997}
Sepp Hochreiter and J{\"u}rgen Schmidhuber.
\newblock Long {{Short-Term Memory}}.
\newblock \emph{Neural Computation}, 9\penalty0 (8):\penalty0 1735--1780,
  November 1997.
\newblock ISSN 0899-7667.
\newblock \doi{10.1162/neco.1997.9.8.1735}.

\bibitem[Ardila et~al.(2020)Ardila, Branson, Davis, Henretty, Kohler, Meyer,
  Morais, Saunders, Tyers, and Weber]{Ardila2020}
Rosana Ardila, Megan Branson, Kelly Davis, Michael Henretty, Michael Kohler,
  Josh Meyer, Reuben Morais, Lindsay Saunders, Francis~M Tyers, and Gregor
  Weber.
\newblock Common voice: {{A}} massively-multilingual speech corpus.
\newblock In \emph{{{LREC}} 2020 - 12th {{International Conference}} on
  {{Language Resources}} and {{Evaluation}}, {{Conference Proceedings}}}, pages
  4218--4222, 2020.
\newblock ISBN 979-10-95546-34-4.

\bibitem[Skirg{\aa}rd et~al.(2017)Skirg{\aa}rd, Roberts, and
  Yencken]{Skirgard2017}
Hedvig Skirg{\aa}rd, Sean~G Roberts, and Lars Yencken.
\newblock Why are some languages confused for others? {{Investigating}} data
  from the great language game.
\newblock \emph{PLoS ONE}, 12\penalty0 (4), 2017.
\newblock ISSN 19326203.
\newblock \doi{10.1371/journal.pone.0165934}.

\bibitem[Fant et~al.(2000)Fant, Kruckenberg, and Liljencrants]{Fant2000}
Gunnar Fant, Anita Kruckenberg, and Johan Liljencrants.
\newblock Acoustic-phonetic {{Analysis}} of {{Prominence}} in {{Swedish}}.
\newblock In \emph{Intonation}, pages 55--86. {Springer, Dordrecht}, 2000.
\newblock \doi{10.1007/978-94-011-4317-2_3}.

\bibitem[Sluijter and Van~Heuven(1996)]{Sluijter1996}
Agaath M~C Sluijter and Vincent~J Van~Heuven.
\newblock Spectral balance as an acoustic correlate of linguistic stress.
\newblock Technical report, 1996.

\bibitem[Jadoul et~al.(2018)Jadoul, Thompson, and {de Boer}]{Jadoul2018}
Yannick Jadoul, Bill Thompson, and Bart {de Boer}.
\newblock Introducing {{Parselmouth}}: {{A Python}} interface to {{Praat}}.
\newblock \emph{Journal of Phonetics}, 71:\penalty0 1--15, November 2018.
\newblock ISSN 00954470.
\newblock \doi{10.1016/j.wocn.2018.07.001}.

\bibitem[Boersma and Weenink(2018)]{Boersma2018}
Paul Boersma and David Weenink.
\newblock Praat: Doing phonetics by computer, 2018.

\bibitem[Gers et~al.(2000)Gers, Schmidhuber, and Cummins]{Gers2000}
Felix Gers, J{\"u}rgen Schmidhuber, and Fred Cummins.
\newblock Learning to {{Forget}}: {{Continual Prediction}} with {{LSTM}}.
\newblock \emph{Neural computation}, 12:\penalty0 2451--71, October 2000.
\newblock \doi{10.1162/089976600300015015}.

\bibitem[Abadi et~al.(2016)Abadi, Barham, Chen, Chen, Davis, Dean, Devin,
  Ghemawat, Irving, Isard, et~al.]{Abadi2016}
Mart{\'\i}n Abadi, Paul Barham, Jianmin Chen, Zhifeng Chen, Andy Davis, Jeffrey
  Dean, Matthieu Devin, Sanjay Ghemawat, Geoffrey Irving, Michael Isard, et~al.
\newblock {{TensorFlow}}: {{Large-Scale Machine Learning}} on {{Heterogeneous
  Distributed Systems}}, March 2016.

\bibitem[Srivastava et~al.(2014)Srivastava, Hinton, Krizhevsky, Sutskever, and
  Salakhutdinov]{Srivastava2014}
Nitish Srivastava, Geoffrey Hinton, Alex Krizhevsky, Ilya Sutskever, and Ruslan
  Salakhutdinov.
\newblock Dropout: {{A Simple Way}} to {{Prevent Neural Networks}} from
  {{Overfitting}}.
\newblock \emph{Journal of Machine Learning Research}, 15\penalty0
  (56):\penalty0 1929--1958, 2014.
\newblock ISSN 1533-7928.

\bibitem[Semeniuta et~al.(2016)Semeniuta, Severyn, and Barth]{Semeniuta2016}
Stanislau Semeniuta, Aliaksei Severyn, and Erhardt Barth.
\newblock Recurrent dropout without memory loss.
\newblock In \emph{{{COLING}} 2016 - 26th {{International Conference}} on
  {{Computational Linguistics}}, {{Proceedings}} of {{COLING}} 2016:
  {{Technical Papers}}}, pages 1757--1766, 2016.
\newblock ISBN 978-4-87974-702-0.
\newblock \doi{10.5281/zenodo.546212}.

\bibitem[{van der Maaten} and Hinton(2008)]{VanderMaaten2008}
Laurens {van der Maaten} and Geoffrey Hinton.
\newblock Visualizing {{Data}} using t-{{SNE}}.
\newblock \emph{Journal of Machine Learning Research}, 1:\penalty0 1--48, 2008.

\bibitem[Pedregosa et~al.(2011)Pedregosa, Varoquaux, Gramfort, Michel, Thirion,
  Grisel, Blondel, Prettenhofer, Weiss, Dubourg, Vanderplas, Passos,
  Cournapeau, Brucher, Perrot, and Duchesnay]{scikit-learn}
F~Pedregosa, G~Varoquaux, A~Gramfort, V~Michel, B~Thirion, O~Grisel, M~Blondel,
  P~Prettenhofer, R~Weiss, V~Dubourg, J~Vanderplas, A~Passos, D~Cournapeau,
  M~Brucher, M~Perrot, and E~Duchesnay.
\newblock Scikit-learn: {{Machine Learning}} in {{Python}}.
\newblock \emph{Journal of Machine Learning Research}, 12:\penalty0 2825--2830,
  2011.

\bibitem[Wattenberg et~al.(2016)Wattenberg, Vi{\'e}gas, and
  Johnson]{Wattenberg2016}
Martin Wattenberg, Fernanda Vi{\'e}gas, and Ian Johnson.
\newblock How to {{Use}} t-{{SNE Effectively}}.
\newblock \emph{Distill}, 1\penalty0 (10):\penalty0 e2, October 2016.
\newblock ISSN 2476-0757.
\newblock \doi{10.23915/distill.00002}.

\bibitem[Gal and Ghahramani(2016)]{Gal2016}
Yarin Gal and Zoubin Ghahramani.
\newblock Dropout as a {{Bayesian Approximation}}.
\newblock In \emph{33rd {{International Conference}} on {{Machine Learning}},
  {{ICML}} 2016}, volume~3, pages 1661--1680, 2016.
\newblock ISBN 978-1-5108-2900-8.

\bibitem[Dellwo(2006)]{Dellwo2006}
V~Dellwo.
\newblock Rhythm and {{Speech Rate}}: {{A Variation Coefficient}} for
  {{deltaC}}.
\newblock In \emph{Language and Language-Processing}, number 1999, pages
  231--241. 2006.
\newblock ISBN 3-631-50311-3.

\bibitem[Zou and Hastie(2005)]{Zou2005}
Hui Zou and Trevor Hastie.
\newblock Regularization and variable selection via the elastic net.
\newblock \emph{J. R. Statist. Soc. B}, 67\penalty0 (2):\penalty0 301--320,
  2005.

\bibitem[Ramus(2002)]{Ramus2002}
Franck Ramus.
\newblock Acoustic correlates of linguistic rhythm: {{Perspectives}}.
\newblock In \emph{Proceedings of {{Speech Prosody}} 2002}, pages 115--120,
  2002.
\newblock \doi{10.1.1.16.326}.

\bibitem[Loukina et~al.(2011)Loukina, Kochanski, Rosner, Keane, and
  Shih]{Loukina2011}
Anastassia Loukina, Greg Kochanski, Burton Rosner, Elinor Keane, and Chilin
  Shih.
\newblock Rhythm measures and dimensions of durational variation in speech.
\newblock \emph{The Journal of the Acoustical Society of America}, 129\penalty0
  (5):\penalty0 3258--3270, 2011.
\newblock ISSN 0001-4966.
\newblock \doi{10.1121/1.3559709}.

\bibitem[Dellwo(2007)]{Dellwo2007}
Volker Dellwo.
\newblock Czech {{Speech Rhythm}} and the {{Rhythm Class Hypothesis}}.
\newblock \emph{English}, \penalty0 (August):\penalty0 1241--1244, 2007.

\bibitem[Nespor(1990)]{Nespor1990}
Marina Nespor.
\newblock On the rhythm parameter in phonology.
\newblock In \emph{Logical {{Issues}} in {{Language Acquisition}}}, pages
  157--176. {De Gruyter Mouton}, {Berlin, Boston}, 1990.
\newblock ISBN 978-3-11-087037-4.
\newblock \doi{10.1515/9783110870374-009}.

\bibitem[Mairano(2011)]{Mairano2011}
Paolo Mairano.
\newblock \emph{Rhythm Typology: Acoustic and Perceptive Studies}.
\newblock PhD thesis, Universita degli studi di Torino, March 2011.

\bibitem[Shannon et~al.(1995)Shannon, Zeng, Kamath, Wygonski, and
  Ekelid]{Shannon1995}
Robert~V Shannon, Fan~Gang Zeng, Vivek Kamath, John Wygonski, and Michael
  Ekelid.
\newblock Speech recognition with primarily temporal cues.
\newblock \emph{Science}, 270\penalty0 (5234):\penalty0 303--304, 1995.
\newblock ISSN 00368075.
\newblock \doi{10.1126/science.270.5234.303}.

\bibitem[Kohler(2009{\natexlab{b}})]{Kohler2009}
Klaus~J. Kohler.
\newblock Whither speech rhythm research?
\newblock \emph{Phonetica}, 66\penalty0 (1-2):\penalty0 5--14,
  2009{\natexlab{b}}.
\newblock ISSN 00318388.
\newblock \doi{10.1159/000208927}.

\bibitem[Beaufays et~al.(2014)Beaufays, Sak, and Senior]{Beaufays2014}
Francoise Beaufays, Hasim Sak, and Andrew Senior.
\newblock Long {{Short-Term Memory Recurrent Neural Network Architectures}} for
  {{Large Scale Acoustic Modeling Has}}.
\newblock \emph{Interspeech}, \penalty0 (September):\penalty0 338--342, 2014.
\newblock ISSN 0028-0836.
\newblock \doi{arXiv:1402.1128}.

\bibitem[Karpathy et~al.(2016)Karpathy, Johnson, and {Fei-Fei}]{Karpathy2015}
Andrej; Karpathy, Justin; Johnson, and Li~{Fei-Fei}.
\newblock Visualizing and understanding recurrent networks.
\newblock In \emph{{{ICLR Worshop}}}, June 2016.
\newblock ISBN 978-3-319-10589-5.
\newblock \doi{10.1007/978-3-319-10590-1_53}.

\bibitem[Cichy and Kaiser(2019)]{Cichy2019}
Radoslaw~M. Cichy and Daniel Kaiser.
\newblock Deep {{Neural Networks}} as {{Scientific Models}}.
\newblock \emph{Trends in Cognitive Sciences}, 23\penalty0 (4):\penalty0
  305--317, April 2019.
\newblock ISSN 1879307X.
\newblock \doi{10.1016/j.tics.2019.01.009}.

\bibitem[Cammarata et~al.(2020)Cammarata, Carter, Goh, Olah, Petrov, and
  Schubert]{Cammarata2020}
Nick Cammarata, Shan Carter, Gabriel Goh, Chris Olah, Michael Petrov, and
  Ludwig Schubert.
\newblock Thread: {{Circuits}}.
\newblock \emph{Distill}, March 2020.
\newblock ISSN 2476-0757.
\newblock \doi{10.23915/distill.00024}.

\bibitem[Ramus and Mehler(1999)]{ramus1999c}
F.~Ramus and J.~Mehler.
\newblock Language identification with suprasegmental cues: A study based on
  speech resynthesis.
\newblock \emph{The Journal of the Acoustical Society of America}, 105\penalty0
  (1):\penalty0 512--521, January 1999.
\newblock ISSN 0001-4966.
\newblock \doi{10.1121/1.424522}.

\bibitem[Kubilius et~al.(2016)Kubilius, Bracci, and {Op de
  Beeck}]{Kubilius2016}
Jonas Kubilius, Stefania Bracci, and Hans~P. {Op de Beeck}.
\newblock Deep {{Neural Networks}} as a {{Computational Model}} for {{Human
  Shape Sensitivity}}.
\newblock \emph{PLoS Computational Biology}, 12\penalty0 (4):\penalty0
  e1004896, April 2016.
\newblock ISSN 15537358.
\newblock \doi{10.1371/journal.pcbi.1004896}.

\bibitem[Wichmann and Geirhos(2023)]{Wichmann2023}
Felix~A. Wichmann and Robert Geirhos.
\newblock Are {{Deep Neural Networks Adequate Behavioral Models}} of {{Human
  Visual Perception}}?
\newblock \emph{Annual Review of Vision Science}, 9\penalty0 (1):\penalty0
  501--524, 2023.
\newblock \doi{10.1146/annurev-vision-120522-031739}.

\bibitem[Leavitt and Morcos(2020)]{Leavitt2020}
Matthew~L Leavitt and Ari~S Morcos.
\newblock Towards falsifiable interpretability research.
\newblock In \emph{{{NeurIPS}} 2020, {{ML-RSA Workshop}}}, 2020.

\end{thebibliography}

\end{document}